# Spreading the gospel: The Bohr atom popularised

Helge Kragh and Kristian Hvidtfelt Nielsen[*]

Centre for Science Studies, Aarhus University, Aarhus, Denmark

## Summary

The emergence of quantum theory in the early decades of the twentieth century was accompanied by a wide range of popular science books, all of which presented in words and in images new scientific ideas about the structure of the atom. The work of physicists such as Ernest Rutherford and Niels Bohr, among others, was pivotal to the so-called planetary model of the atom, which, still today, is used in popular accounts and in science textbooks. In an attempt to add to our knowledge about the popular trajectory of the new atomic physics, this paper examines one book in particular, coauthored by Danish science writer Helge Holst and Dutch physicist and close collaborator of Niels Bohr, Hendrik A. Kramers. Translated from Danish into four European languages, the book not only presented contemporary ideas about the quantum atom, but also went into rather lengthy discussions about unresolved problems. Moreover, the book was quite explicit in identifying the quantum atom with the atom as described by Bohr's theory. We argue that Kramers and Holst's book, along with other atomic books, was a useful tool for physicists and science popularisers as they grappled with the new quantum physics.

## Contents

---

[*] E-mails: : helge.kragh@ivs.au.dk and khn@ivs.au.dk

## 1. Introduction

The early twentieth century saw the rise of the 'new physics' of relativity and quantum theory and, across Europe and the USA, its popularisation through numerous books, lectures, and articles aimed at a general audience. Although never as widely disseminated and discussed in public as the relativity theory of Albert Einstein, the new quantum theory of the atom, first proposed by Niels Bohr in 1913, won popular renown in the late 1910s and early 1920s through the work of physicists and science popularisers. As observed by historians such as Michael Whitworth and Peter Bowler, the efforts ranged widely in terms of 'popularity': some were genuinely popular in terms of exposition, marketing, and readership; others were quite technical and generally obscure to, and probably never intended to reach, wider audiences.[1] Many prominent scientists, including Einstein and Bohr, were actively engaged in the popularisation of the new physics; others, such as Werner Heisenberg and Paul Dirac, restricted their interest to the elaboration of new scientific knowledge.[2] As a result of the wide spectrum of popular science and the sustained interest in popular science on behalf of some scientists, the boundary between 'real physics' and 'popular physics' never was entirely clear-cut. The fact that the new physics, partly because of its counterintuitive results and partly because of the high degree of complicated mathematics involved, has played an important role in distancing scientific knowledge from lay opinion should not lead us to consider the gap between real and popular science as given. The historical genealogy of this gap provides us with important insights into foundational issues pertaining to science and the public.[3]

---

[1] Michael Whitworth, 'The Clothbound Universe: Popular Physics Books, 1919-39', *Publishing History*, 40 (1996), 53-82. Peter J. Bowler, *Science for All: The Popularization of Science in Early Twentieth-Century Britain* (Chicago: University of Chicago Press, 2009).

[2] As to Einstein, he was keenly interested in popularising his theory of relativity, as shown, for example, by his early semipopular presentation *Über die Spezielle und die allgemeine Relativitätstheorie* (Braunschweig: Vieweg & Sohn, 1917), translated into English as *Relativity: The Special and General Theory* (New York: Hartsdale House, 1920). Although Einstein had spared no pains in his 'endeavour to present the main ideas in the simplest and most intelligible form', the book presumed 'a standard of education corresponding to that of a university matriculation examination' (preface, p. v). On Einstein as a popular science writer, see Elke Flatau, 'Albert Einstein als wissenschaftlicher Autor', Max Planck Institute for the History of Science, Preprint 293 (2005). On Werner Heisenberg, see David C. Cassidy, *Uncertainty: The Life and Science of Werner Heisenberg* (New York: W. H. Freeman, 1992); on Paul Dirac, see Helge Kragh, *Dirac: A Scientific Biography* (Cambridge: Cambridge University Press, 1991) and Graham Farmelo, *The Strangest Man: The Hidden Life of Paul Dirac, Quantum Genius* (London: Faber and Faber, 2009).

The historiography of popular science considers popular science as integral, even fundamental, to the history of science proper.[4] The historical interpretation of scientific developments needs to take into account the distinction between, on the one hand, scientific knowledge and scientific communications, and on the other hand, other kinds of knowledge and other modes of communication. To be sure, for several reasons, the new physics of the twentieth century marks an important shift in the history of emerging demarcations between science and popular science. Scientists and others saw relativity and quantum theory, along with non-Euclidian geometry, as emblematic for a sharp cognitive divide between scientific knowledge and public opinion.[5] The new physics and mathematics made it clear to everyone that scientific knowledge was difficult to access, bordering on the incomprehensible. Whereas, previously, popular science could be seen as an extension of scientific epistemology in public domains, the new physics required translating sophisticated mathematics and highly technical language into everyday language and simple cognitive models, such as images of the atom as a planetary system. Moreover, new venues of mass-communication, new techniques of mass-production of texts and images, and the ever-increasing demand for popular science (probably enforced by the grand claims of the new physics) also contributed to the 'boom' of popular science during the early decades of the twentieth century.[6] Finally, as Andreas Daum notes, the enactment of a gap between science and popular science not only served to denigrate popular science. On the contrary, scientists and popularisers often have appealed to this gap in order to make a positive case for popular science, i.e., 'to present popular knowledge as something positive and necessary'.[7]

---

[3] Bernadette Bensaude-Vincent, 'A Genealogy of the Increasing Gap between Science and the Public', *Public Understanding of Science*, 10 (2001), 99-113.

[4] Jonathan R. Topham et al., 'Focus: Historizicing "Popular Science"', *Isis*, 100 (2009), 310-368. For a historiographical discussion of popular science and its relation to 'history of science proper', see Roger Cooter and Stephen Pumfrey, 'Separate Spheres and Public Places: Reflections on the History of Science Popularization and Science in Popular Culture', *History of Science*, 32 (1994), 237-267, who deplore 'the reluctance of historians of all kinds to commit themselves to inquiry into popular science' (p. 246).

[5] Bensaude-Vincent (note 3), pp. 105-108.

At the core of this paper is an attempt to engage historically with the establishment of the boundary between 'real' and 'popular' atomic theory in the period from about 1915 to 1925. During this decade atomic and quantum physics was dominated by the theory of atomic structure proposed by Niels Bohr in 1913, a theory which also attracted a good deal of public attention. Foremost among the popular expositions of Bohr's theory was a book published by Helge Holst and Hendrik A. Kramers in 1922, which book we analyse in some detail. Following a section on the early reception of Bohr's theory in scientific circles we survey the field of 'atomic books', that is, popular books on atomic theory, thereby providing insights into the kind of popular books written to present the new scientific models of the atom to wider audiences. We largely restrict our survey to books published in Germany and England. The book authored by Holst and Kramers was a rare, joint effort between a Danish science populariser and one of Bohr's close collaborators. Aimed at the general reader with an interest in contemporary physics, the book was received as a welcome addition to the many popular accounts of the new quantum theory of the atom, but also as a vehicle for 'spreading the gospel' of Bohr's interpretation of quantum theory, parts of which were still being debated by atomic physicists.[8]

[6] Peter Bowler, cited in Scott Keir, 'The Booms of Popular Science', *Nature.com Blogs*, <http://blogs.nature.com/scottkeir/2010/04/03/the-booms-of-popular-science> [accessed 4 September 2011].

[7] Andreas W. Daum, 'Varieties of Popular Science and the Transformations of Public Knowledge', *Isis*, 100 (2009), 319-332, on p. 320.

 The Kramers-Holst book, with its explicit omission of mathematical reasoning and its introduction to varieties of atomic theorising throughout the ages, effectively established a boundary between real physics (based on mathematics and experimentation) and popular physics (written in clear and simple language, and supported by a historical narrative). The authors clearly saw popular physics as an important venue for presenting fundamental ideas and significant results, but also as an opportunity for discussing key problems and addressing the physical meaning of the Bohr theory. They wanted, in short, the book to serve as 'a stimulus to further study of the Bohr theory'.[9]

---

[8] Historical comments on the Kramers-Holst book include Max Dresden, *H. A. Kramers: Between Tradition and Revolution* (Berlin: Springer, 1987), pp. 132-134, and Arne Schirrmacher, 'Bohrsche Bahnen in Europa: Bilder und Modelle zur Vermittlung des Modernen Atom', in Charlotte Bigg and Jochen Hennig, eds, *Atombilder: Ikonographie des Atoms in Wissenchaft und Öffentlichkeit des 20. Jahrhunderts* (Munich: Deutsches Museum, 2009). Dresden finds it 'not inappropriate to describe the book by Kramers and Holst as a truly missionary venture to spread the gospel according to Bohr' (p. 134). As we shall see, although it can indeed be described as missionary, the authors did not hide the provisional and incomplete nature of Bohr's theory.

[9] H. A. Kramers and H. Holst, *The Atom and the Bohr Theory of its Structure* (London: Gyldendal, 1923), p. ix.

## 2. The Bohr atom and its reception

Financed by a stipend from the Carlsberg Foundation, young Bohr spent the period from September 1911 to July 1912 in England, first in Cambridge with J. J. Thomson and subsequently in Manchester with Ernest Rutherford. It was during his stay in Manchester that he came upon the idea of combining Rutherford's new hypothesis of the nuclear atom with Planck's quantum theory, an idea he developed into a full-blown atomic theory in a sequel of three seminal papers that appeared in the summer and fall of 1913. Published in the *Philosophical Magazine*, he presented the well-known planetary model of the atom that can still be met in elementary textbooks in physics and chemistry.[10]

What is of interest in the present context is merely that Bohr pictured the atom as consisting of a tiny positive nucleus surrounded by one or more electrons moving in definite, so-called stationary orbits around it. This may sound uncontroversial, much like the planets orbiting the sun, but according to the authoritative theory of classical electrodynamics a revolving electron emits electromagnetic energy, causing an atom of the Bohr-Rutherford type to collapse. To avoid the radiation catastrophe, Bohr *postulated* that electrons in stationary orbits do not obey the laws of electrodynamics. Moreover, he assumed that electrons in higher ('excited') energy states will spontaneously 'jump' from the higher to a lower stationary state, by which process the atom will emit a discrete amount of radiation. The frequency of the emitted light is given by $\Delta E/h$, where $\Delta E$ is the energy difference between the two states and $h$ is the quantum constant introduced by Planck in 1900. Only when the atom is in the lowest possible energy state, known as the ground state, will it be stable and not emit radiation.

---

[10] Niels Bohr, 'On the Constitution of Atoms and Molecules', *Philosophical Magazine*, 26 (1913), 1-25; 476-502; 857-875. There are several historical analyses of Bohr's atomic theory, see for example Jagdish Mehra and Helmut Rechenberg, *The Historical Development of Quantum Theory*, vol. 1 (New York: Springer, 1982) and Olivier Darrigol, *From c-Numbers to q-Numbers: The Classical Analogy in the History of Quantum Theory* (Berkeley: University of California Press, 1992). For a more accessible review and a full biography of Bohr, see Abraham Pais, *Niels Bohr's Times, in Physics, Philosophy, and Polity* (Oxford: Clarendon Press, 1991), pp. 176-209.

Based on the two postulates or assumptions, and supplied with some further hypotheses that lacked independent justification, Bohr succeeded in accounting quantitatively for the line spectrum of hydrogen that until then had defied explanation. In addition, he calculated the ionisation energy (the energy it takes to detach the electron from the atom) in agreement with experiment and predicted several other phenomena that were soon verified experimentally. In short, Bohr's theory was greatly successful from an empirical point of view. Its impressive explanatory and predictive power convinced many physicists to take it seriously in spite of its doubtful foundation in the two postulates. Many of those who adopted the theory used it selectively and opportunistically: while they accepted the physical model, they either denied or ignored its theoretical foundation. As James Jeans pointedly said at the meeting of the British Association for the Advancement of Science in September 1913, 'The only justification … for these assumptions is the very weighty one of success'.[11]

In spite of initial scepticism and scattered opposition, Bohr's theory was generally if somewhat hesitatingly welcomed by a large part of the physics community. The conversion to the Bohr atom did not occur instantly, but latest by the summer of 1915 the theory was well known and accepted as superior to alternative conceptions of the constitution of the atom. It first attracted interest among English physicists, among whom Rutherford, Henry Moseley, and Owen Richardson supported it from an early date, while John Nicholson and some other physicists resisted the new theory.[12]

---

[11] James Jeans, 'Discussion on Radiation', *Report, British Association for the Advancement of Science* (London: J. Murray, 1914), 376-386, on p. 379. This was the meeting at which Bohr's theory was first discussed in public and also the occasion for the first mention of it in the press. Reporting from the meeting, *The Times* of London briefly mentioned on 13 September Jeans's account of 'Dr. Bohr's ingenious explanation of the hydrogen spectrum'.

German physicists were slower to respond to Bohr's atom, but when they did, they did it effectively and with great consequences. Contrary to their colleagues in England, they developed the theory scientifically and turned it into a more general and even more powerful theory. This important development, which took place during the difficult war years, was primarily the work of Arnold Sommerfeld and his innovative school in Munich. The result was that by 1918 the Bohr (or Bohr-Sommerfeld) theory was much better known in Germany than in England. The further development of Bohr's atomic theory, largely identical to what is known as the 'old quantum theory' (as distinct from quantum mechanics), was also dominated by German physicists. Until the fall of the theory in 1925, the main centres of quantum and atomic theory were Copenhagen, Munich, and Göttingen. It should be noted that the Bohr atom changed considerably over time. Whereas Bohr originally conceived atoms as planar configurations of rings populated with evenly spaced electrons, by 1922 the picture had changed to a more complicated three-dimensional model in which the electrons moved in elliptic orbits of different eccentricities and spatial orientations.

Bohr's atomic theory never attracted the same kind of public attention as, for example, Einstein's theory of relativity. Yet it was known at an early stage not only by physicists interested in the structure of matter but also by a broader segment of scientists and lay readers of the general science literature. Journals such as *Nature* (England), *Science* (the USA), and *Die Naturwissenschaften* (Germany) were not only read by professional scientists but also by many people with a general interest in the sciences.

---

[12] On the British opposition, and the one of Nicholson in particular, see Helge Kragh, 'Resisting the Bohr Atom: The Early British Opposition', *Physics in Perspective*, 13 (2011), 4-35.

While the early discussion of the Bohr atom in *Nature* was of a technical nature and of interest mostly to physicists, readers of *Science*, the journal of the American Association of the Advancement of Science, were more broadly informed. For example, the July 1914 issue of the journal included a survey article by Arthur S. Eve based on a meeting of the Royal Society of Canada on the structure of the atom. Eve, a former assistant (and later biographer) of Rutherford and since 1903 professor of physics in Montreal, presented the ideas of the 'brilliant young Dane, Bohr' whose work 'is remarkable as leading to excellent numerical verification'.[13] Half a year later *Science* brought another survey article which praised the Bohr-Rutherford model of the atom as a great advance, even one that 'will probably remain, suffering but little change in the future'.[14] The author, G. Walter Stewart of the University of Iowa City, recognised the weaknesses of the model but did not find it damaging that Bohr's theory had difficulties with the more complex atoms: 'When one contemplates the narrow scope of even this brilliant theory, what a limitless field for research seems ahead!'

The scientifically interested public in German-speaking Europe might get a thorough introduction to the Bohr theory in the pages of the recently founded *Die Naturwissenschaften*, which in March 1914 carried a long article on the subject by Rudolf Seeliger, a young physicist at the Physikalisch-Technische Reichsanstalt in Berlin. Although Seeliger was alert that the theory rested on a problematic foundation and might even not be consistent, his review was generally positive. Like

---

[13] Arthur S. Eve, 'Modern Views on the Constitution of the Atom', *Science*, 40 (1914), 115-121.
[14] G. Walter Stewart, 'The Content and Structure of the Atom', *Science*, 40 (1914), 661-663.

Stewart and several other physicists, he concluded that the empirical strength of the theory overshadowed the conceptual problems associated with it: 'Even though we may be sceptical with respect to the details, I think we have in Bohr's considerations an important and fundamental advance in the knowledge of the origin of spectral lines and series.'[15] Contrary to the reviews in *Science*, the one of Seeliger was of a detailed and rather technical nature, undoubtedly appealing more to physicists than to lay readers.

While magazines like *Science* and *Naturwissenschaften* primarily were directed to scientists and scientifically-competent citizens, at least in one case Bohr's theory also appeared early on in a genuinely popular science journal. The US *Popular Science Monthly*, founded in 1872 by Edward Youmann, included in the summer of 1915 a lecture that Rutherford had given to the National Academy of Sciences in Washington D.C. in April the previous year. In this detailed yet non-technical review Rutherford covered the most recent developments in atomic and subatomic physics, including not only the nuclear atom but also isotopes, Wilson's cloud chamber, and Moseley's measurements of the characteristic X-ray lines from elements. The question of the spectra, he said, 'has been attacked in a series of remarkable papers by Bohr, who concludes that the complexity of the spectrum is not due to the complexity of the atom but to the variety of modes in which an electron can emit radiation'. Although Rutherford expressed confidence in Bohr's model, at the same time he admitted that 'there is room for much difference of opinion as to the interpretation of the rather revolutionary assumptions made by Bohr'.[16]

---

[15] Rudolf Seeliger, 'Moderne Anschauungen über die Entstehung der Spektrallinien und der Serienspektren', *Die Naturwissenschaften*, 2 (1914), 285-290, 309-314, on p. 313. *Die Naturwissenschaften*, founded in 1913 and published by the Springer Verlag, was until 1935 edited by the physicist Arnold Berliner. Associated with the Kaiser Wilhelm Gesellschaft, the magazine was subtitled *Wochenschrift für die Fortschritte der Naturwissenschaften, der Medizin und der Technik* (Weekly Publication for the Advances in the Natural Sciences, Medicine and Technology). It was in many ways the German equivalent to the British *Nature* and the US *Science*.

As one might expect, Bohr's work attracted early interest in Denmark even though the atomic theory was a product of Manchester rather than Copenhagen. We have an interesting example in the 25-volume encyclopaedia *Salmonsens Konversations Leksikon*, which in its second edition of 1915 included an entry on Niels Bohr (and also one on his father, the physiologist Christian Bohr, and his younger brother, the mathematician Harald Bohr). The entry mentioned his early work on the electron theory of metals, and also his recent work on atomic theory:

> Based on Rutherford's atomic model, but supplemented by a few remarkable and revolutionary hypotheses, he explains many of the physical, and especially optical, properties of matter (particularly in the case of hydrogen, the simplest element). Thus the atomic model yields directly and quantitatively correctly the spectrum of hydrogen. The considerations which are established in this way are expected to be greatly important for the further research in these areas.[17]

---

[16] Ernest Rutherford, 'The Constitution of Matter and the Evolution of the Elements', *Popular Science Monthly*, 87 (August 1915), 104-142, on pp. 139-140. The magazine was at the time about to change its profile. While the existing version was still of a rather scholarly nature, often with extensive articles written by recognised scientists, by the end of 1915 the journal changed to a format with numerous small and easy to read articles written by its reporters. At the same time the number of illustrations grew dramatically. The magazine became popular in a sense different from the older one, such as we discuss in our introduction.

[17] Carl Blangstrup, ed., *Salmonsens Konversations Leksikon*, vol. 3 (Copenhagen: J. H. Schultz, 1915). The author was the physicist Hans Marius Hansen, a close friend of Bohr.

In the same year Bohr's theory made its entry in the standard textbook used by physics students at the University of Copenhagen, a revised fourth edition of Christian Christiansen's textbook first published 1892-1894.[18]

### 3. Popular expositions of atomic theory

It was only after the end of the First World War that Bohr's atomic theory won general recognition and became the unquestioned basis for research in atomic and molecular structure such as communicated in the physics journals. The first books of a non-specialist nature expounding the theory date from the years around 1920. The first was perhaps *Die Atomtheorie* from 1918, a small book based on a series of lectures given by Leo Graetz, professor of physics in Munich. Introducing the Bohr atom as a 'an entire solar system' with the electrons whirling around the nucleus 'like the earth moves around the sun', Graetz emphasised the important difference that the electrons, contrary to the planets, could only move in discrete orbits given by a whole number. Remarkably, contrary to all previous pictures of the atom the quantum atom did not have a definite volume. As Bohr had pointed out in his 1913 trilogy, the size of an atom in an excited state increased with the associated quantum number. A highly excited atom might have a size, say, a radius of 0.01 mm, that made it easily visible in a microscope! As Graetz pointed out, the new atomic theory was based on 'arbitrary assumptions', namely the two quantum postulates that could only be justified by their empirical consequences in spectroscopy.[19] On the other hand, these consequences had been verified so convincingly that there could be no doubt of the essential truth of the theory. Graetz illustrated his book with several pictures of atoms and molecules, representing the electrons as much larger bodies than the atomic nucleus (Figure 1).[20]

---

[18] Christian Christiansen, *Lærebog i Fysik* (Copenhagen: Gyldendal, 1915), p. 456, revised by Martin Knudsen, the successor of Christiansen as professor of physics in Copenhagen.

[19] Leo Graetz, *Die Atomtheorie in ihrer neuesten Entwickelung* (Stuttgart: J. Engelhorns Nachf., 1918), p. 78. The book must have sold well, for in 1922 it came out in a fourth printing and the same year it appeared in a Russian translation. Graetz justified his publication by the general interest in atomic theory he had experienced 'not only from physicists and chemists,

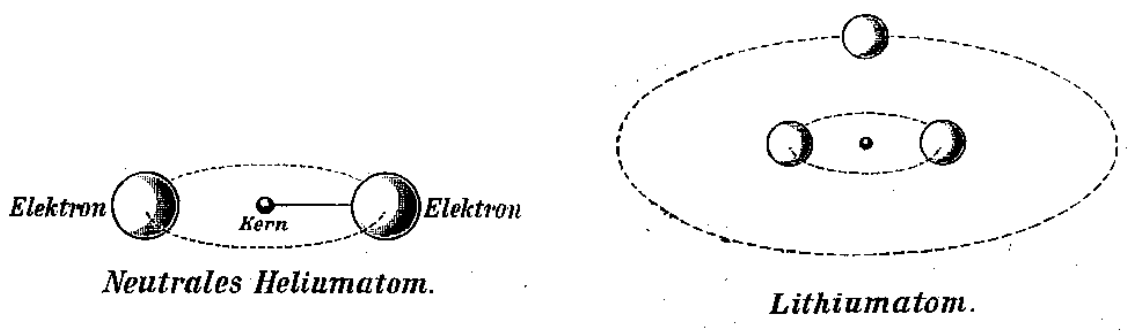

Fig. 1. The Bohr model of helium and lithium according to Leo Graetz's *Atomtheorie* of 1918. The nucleus is depicted as much smaller than the electrons, indicating that Graetz conceived the charged particles as electromagnetic in nature. Source: Graetz 1918 (note 19).

German lay readers could get an updated and more detailed exposition of the Bohr atom from *Die Entwicklung der Atomtheorie*, a book of 1922 written by the physics teacher Paul Kirchberger. About half of this historically organised and careful discussion of atomism dealt with the composite atom of the twentieth century, including a 40-page chapter on the Bohr model which was expounded in considerable detail. Like Graetz and several other authors, Kirchberger pointed out that Bohr's postulates were 'somewhat arbitrary' and that his model of the atom was but also from most scientifically educated laypersons'. Preface dated August 1918. Of the six lectures included in the book, the two last (pp. 60-88) were largely devoted to Bohr's atomic theory.

[20] This was a remnant of the 'electromagnetic world view' according to which the mass of a charged spherical particle was of electromagnetic origin, varying as the square of the charge over the radius. It follows that the radius of the electron must be nearly 2000 times as great as that of the proton. Bohr did not accept this interpretation and never spoke of the size of electrons.

far from visualisable (*anschaulich*) – which may have been the reason why he refrained from including pictures of the atom. Visualisable or not, to Kirchberger it was obvious that Bohr's theory was correct and the only possible way to get further insight in the strange world of the atoms.[21]

By far the most influential book on the new quantum theory of atoms was Sommerfeld's *Atombau und Spektrallinien*, the first edition of which was published in 1919 and which until 1925 appeared in four editions of increasing length. An English translation of the third edition was published in 1923. The only reason to mention this famous work, often referred to as the 'Bible' of quantum theory, in the present context is that Sommerfeld himself thought of it as a popular exposition. In a letter to Einstein of June 1918 he told that he had started writing 'a popular book on "Atomic Structure and Spectral Lines", which in its main text is for chemists but the appendices of which are also for physicists'.[22] In the preface to the first edition Sommerfeld emphasised that his book was meant to be accessible to a generally educated readership (*gemeinverständlich*). However, *Atombau* was not a popular book in the sense of being easy to read and understand for readers with no scientific training. 'Popular science', to scientists like Sommerfeld, took on a different meaning: it provided a venue of communication in which basic ideas and physical concepts could take centre stage. Sommerfeld's book was clear, didactic, and made use of only fairly elementary mathematics (by Sommerfeld's standards). It was hardly accessible to a lay public and probably not read by it. Wide readership was not overly important to authors like Sommerfeld who, in times of some scientific controversy about the viability of the new physics and with very few books available on the topic, saw popular science as yet another means of making new physical knowledge comprehensible and accessible.

---

[21] Paul Kirchberger, *Die Entwicklung der Atomtheorie, gemeinverständlich dargestellt* (Karlsruhe: C. F. Müller, 1922), preface dated October 1921. A second revised edition appeared in 1929. Kirchberger also wrote articles on the new atomic theory for the newspapers, for example 'Moderne Atomtheorie' in *Berliner Tageblatt* of 16 August 1922.

[22] Arnold Sommerfeld, *Atombau und Spektrallinien* (Braunschweig: Vieweg & Sohn, 1919). Sommerfeld to Einstein, June 1918, in Michael Eckert and Karl Märker, eds, *Arnold Sommerfeld. Wissenschaftlicher Briefwechsel*, vol. 1 (Berlin: Verlag für Geschichte der Naturwissenschaften und der Technik, 2000), p. 597. While the first edition of *Atombau* had a length of 550 pages, the fourth edition had expanded to 862 pages.

English books on the quantum atom appeared a little later than in Germany[23] and had a somewhat different character. Among the earliest was *The Atom*, a book by the US physicist, inventor and prolific model-builder Albert Cushing Crehore. However, Crehore's book was of a different genre, as its purpose was not to expound currently accepted knowledge about atomic structure but to promote his own unorthodox model of the atom. This he did in great detail, including some heavy doses of mathematics, and on his way he confronted what at the time had become the standard model, the one of Bohr and his German allies. Crehore admitted that Bohr's theory 'has made a very strong appeal to physicists, who with some reservations may be said to have adopted it as their guiding theory', but he did not find its popularity justified.[24] Like other physicists of a conservative inclination, he objected that there was nothing in the Bohr atom that vibrated with the frequency of the emitted light. Crehore's own alternative was to modify the laws of electrodynamics so as to make them comply with the quantum atom, but his alternative was ignored by mainstream physicists. Readers of Crehore's book could find in it a critical if not unfair exposition of the main features of the Bohr atomic model.

---

[23] The first book which referred to Bohr's atomic model was actually English, namely a book on X-rays written by George W. C. Kaye, a physicist at the National Physical Laboratory. G. W. C. Kaye, *X rays: An Introduction to the Study of Röntgen Rays* (London: Longmans, Green and Co., 1914), on p. 18, preface dated February 1914.

[24] Albert C. Crehore, *The Atom* (New York: Van Nostrand, 1920), preface 14 June 1919, on p. 2. His exposition of the Bohr atom appeared on pp. 24-30. For Crehore and his views of atomic structure, see Kragh 2011 (note 12).

Of the British books dealing with the Bohr model of the atom, we shall only refer to four popular books published in 1923.[25] Two of them were extensive reviews written by physicists and mainly addressed to an audience of physicists, chemists and engineers, although educated lay readers might benefit from them as well.[26] Edward Andrade's *The Structure of the Atom* gave a detailed account of atomic physics, both nuclear and extra-nuclear, including a largely qualitative discussion of the Bohr atom and its relation to experiments. Norman Robert Campbell not only covered the same ground and at largely the same level, his book also appeared with the same title (both books were prefaced April 1923). The two books were informative and competent overviews of the state of art in atomic and quantum physics rather than attempts to disseminate the topic to a broad audience. They can be seen as more qualitative and less demanding versions of Sommerfeld's *Atombau*. The third book to be mentioned was of an altogether different character, for once not written by a scientist but by a science writer and journalist.

John W. N. Sullivan had studied mathematics and science, but without taking a degree, before he turned to literature and popular science. As a science writer he made his name with articles from the spring of 1919 on Einstein's general theory of relativity, and he later wrote a number of books on science, literature and culture. In a small and unpretentious book of 1923, entitled *Atoms and Electrons*, he covered the modern developments in atomic theory, paying much attention to the ideas of the structure of atoms due to the 'brilliant young Danish physicist, Niels Bohr'.[27] Among the topics he dealt with in some detail was the new explanation of the periodic system in terms of electronic orbits that Bohr had expounded in his Nobel lecture and at other occasions. Although not a book exclusively about Bohr's atomic theory, nearly a third of it was a competent if naturally condensed account of this theory. Readers would get the impression that modern atomic theory was solely due to Bohr: apart from a brief reference to Sommerfeld, no other physicists were mentioned as contributors to the theory.

[25] According to a *Catalogue of British Scientific and Technical Books* issued by the British Science Guild, in 1925 there were only 3 scientific books on quantum topics (a subgroup under 'Spectra and Molecular Physics') published in England, out of a total of 318 physics books. The corresponding figures for 1921 were 1 and 269. See Raykumari Williamson, *The Making of Physicists* (Bristol: Adam Hilger, 1987), p. 10.
[26] Edward N. da C. Andrade, *The Structure of the Atom* (London: G. Bell and Sons, 1923). N. Robert Campbell, *The Structure of the Atom* (Cambridge: Cambridge University Press, 1923).

*Atoms and Electrons* was explicitly written as a popular physics book. It appeared in Hodder and Stoughton's series 'People's Library', the object of which was 'in some degree to satisfy that ever-increasing demand for knowledge which is one of the happiest characteristics of our time'. Probably to keep the price low (it sold for 2*s* 6*d*), it contained no pictures, which was unusual for a book of its kind. Another popular atom book was *The ABC of Atoms*, written by the famous philosopher, mathematician and author Bertrand Russell, who, at the time, supported himself and his family as a writer of all kinds of popular books.[28] Basically appealing to the same audience as Sullivan's book, Russell's *ABC* was somewhat more demanding. Like most popularisers, Russell sought to illustrate the 'quite unintelligible' features of Bohr's theory by means of analogies. According to Bohr, electrons moved instantaneously and mysteriously from one stationary orbit to another; according to

[27] John W. N. Sullivan, *Atoms and Electrons* (London: Hodder and Stoughton, 1923), on p. 121, reviewed in *Nature*, 113 (1924), 379-380. On Sullivan as a science writer, see Whitworth 1996 (note 1), who also mentions a few other English popular books dealing with atoms, including Oliver Lodge, *Atoms and Rays: An Introduction to Modern Views on Atomic Structure and Radiation* (New York: George H. Doran, 1924).
[28] Bertrand Russell, *The Autobiography of Bertrand Russell*, vol. 2 (London: Allen and Unwin, 1968), p. 152.

Russell, 'An electron is like a man who, when he is insulted, listens at first apparently unmoved, and then suddenly hits out'.[29]

The year 1923 was a good year for popular and semi-popular works on the quantum atom. A few physicists and some science popularisers joined forces in making accessible and comprehensible the new quantum theory. Although their efforts ranged widely in terms of exposition and 'user-friendliness', they all were convinced that quantum theory was important enough to merit popular accessibility and acceptance. Popular quantum theory required translating mathematical reasoning into everyday language and images based on everyday experiences. In order to go a step deeper into the meaning of popular science at the time, we now turn to a book published in English in the very same year as Andrade's, Campbell's, Sullivan's, and Russell's books; it was a translation of a Danish book co-written by a Danish librarian and Dutch physicist. The stage of the book was set in Copenhagen.

### 4. The Copenhagen context: Bohr, Kramers, Holst, Klein

Niels Bohr considered it important to expound his views of quantum and atomic theory not only to the physics community but also to a broader scientific audience. This he did mostly in the form of lectures, both in Denmark and abroad, and some of these lectures were subsequently published. For example, a lecture given before the Danish Physical Society in December 1913 was published in the Danish journal *Fysisk Tidsskrift* in 1914 and several years later in German, English, and French. In October 1921 he delivered an important address to a joint meeting of the Physical Society and the Chemical Society in Copenhagen, and this address was also published internationally. The same was the case with his Nobel lecture in Stockholm on 11 December 1922. However, Bohr felt no need (and had neither time, nor, we believe, much talent) to expose his views of atomic structure to wider audiences either in the form of magazine articles or a popular book. He did however write an entry on 'Atom' for the thirteenth edition of the *Encyclopaedia Britannica*, but when the volume appeared in 1926 much of the content was obsolete because of the quantum-mechanical revolution.[30]

---

[29] Bertrand Russell, *The ABC of Atoms* (London: Kegan Paul, Trench, Trubner & Co., 1927), p. 63. The first edition from the summer of 1923 was sold at 4*s* 6*d* and printed in 3000 copies (Whitworth 1996, note 1). Russell also wrote a popular book on relativity theory, *The ABC of Relativity* (London: Kegan Paul, Trench, Trubner & Co., 1925), the first edition of which was printed in 2000 copies.

The closest Bohr came to writing a popular book on atomic theory was *The Theory of Spectra and Atomic Constitution*, a small book published by Cambridge University Press in 1922, which also appeared in German and French translations.[31] The book was a collection of three lectures, the two Copenhagen lectures mentioned above (from 1913 and 1921, respectively) and a previously published address to the German Physical Society of 1920. The principal object of the book was 'to emphasize certain general views in a freer form than is usual in scientific treatises and text books', for which reason references and footnotes were left out.[32] Bohr, like many other physicists at the same, saw popularisation as a means to make clear and accessible the general outline of contemporary atomic theory, even though, apart from Bohr's concession to reader-friendliness, the book was highly technical and, at times, partly obscure. Yet in a highly laudable review essay in *Nature* the Cambridge physicist Ralph Fowler evaluated it not only as 'a great work' but also as one which expounded Bohr's atomic theory 'in a simple non-mathematical way which should be capable of being followed by anyone who is prepared to accept the mathematical theorems on which the work is necessarily based'.[33] The mathematical intricacy of atomic physics necessitated a new kind of popular science in which mathematical symbols were kept to a minimum in favour of the fundamental physical meaning of the mathematics.

---

[30] Niels Bohr, 'Atom', *Encyclopaedia Britannica*, 13th ed., Suppl., vol. 1 (1926), 262-267. It is interesting, both from the point of view of popularisation and the history of physics, to compare Bohr's article with the famous article on 'Atom' that Maxwell wrote for the 9th edition of *Encyclopaedia Britannica* from 1875 (pp. 36-49). Bohr made various attempts to present modern atomic physics to a general audience, but mostly in the post-1925 period. See Finn Aaserud, ed., *Niels Bohr Collected Works*, vol. 12 (Amsterdam: Elsevier, 2007).

[31] N. Bohr, *The Theory of Spectra and Atomic Constitution*, translated by A. D. Udden (Cambridge: Cambridge University Press, 1922). N. Bohr, *Drei Aufsätze über Spektren und Atombau* (Braunschweig: Vieweg & Sohn, 1922). N. Bohr, *Les spectres et la structure de l'atome: trois conférences*, translated by A. Corvisy (Paris: J. Hermann & Cie., 1923). A second English and German edition, slightly revised and provided with a couple of appendices, came out in 1924.

[32] Bohr 1922 (note 31, English edition), p. vi.

What Bohr did not do, namely, write a comprehensive popular exposition of his atomic theory, was done by his close collaborator, the Dutchman Hendrik Antonie Kramers, and the Danish physics-trained librarian and author Helge Holst. Danish readers might also learn about Bohr's theory from expositions given by other Danish scientists, including the distinguished chemist Niels Bjerrum.[34] Before discussing the product of the Kramers-Holst collaboration, *The Atom and the Bohr Theory of its Structure*, it will be useful to introduce the two authors. While H. A. ('Hans') Kramers is not well known to the public, in the history of science he is recognised as one of the great theoretical physicists of the twentieth century.[35] After

---

[33] Ralph H. Fowler, 'The Structure of the Atom', *Nature*, 111 (1923), 523-525, on p. 523.

[34] Niels Bjerrum, 'Fysik og Kemi', pp. 71-194 in Torsten Brodén, Niels Bjerrum, and Elis Strömgren, *Matematiken og de Eksakte Naturvidenskaber i det Nittende Aarhundrede* (Copenhagen: Gyldendal, 1925). Another example, mostly aimed at engineers, was Edvard S. Johansen, *Moderne Anskuelser om Elektricitet og Stof* (Copenhagen: Gjellerup, 1920).

[35] The standard biography of Kramers is Dresden (note 8), which covers his life and science in impressive details. A briefer account of Kramers's contributions to physics is given by Hendrik Casimir in *Dictionary of Scientific Biography*, vol. 7 (1973), 491-494. For Bohr's appreciation of Kramers, see the memorial address in *Nederlandsch Tijdschrift voor*

having graduated from the University of Leiden under Paul Ehrenfest, in 1916 the 21-year-old Kramers arrived in Copenhagen to become Bohr's first scientific assistant. He remained with Bohr until 1926, when he was appointed professor of theoretical physics at the University of Utrecht. During this fruitful period he collaborated intensely with Bohr, first and foremost on the frustratingly difficult problem of the quantum structure of helium, the simplest atom apart from hydrogen.

On the instigation of Bohr, Kramers wrote in 1919 a doctoral dissertation in which he applied a mathematically sophisticated version of the correspondence principle to calculate the intensity of the spectral lines of hydrogen. This work, by which he made his name in international physics, demonstrated convincingly that Bohr's correspondence principle was a powerful tool in quantitative atomic theory. If Bohr was the philosopher and architect of the correspondence principle (which he was), Kramers was its computer. His most important contribution to the old quantum theory was probably a theory of dispersion of light published in 1924 which he further developed in a paper of 1925 written jointly with the even younger Werner Heisenberg. Kramers's works on dispersion proved important stepping stones for Heisenberg in his construction of quantum mechanics. A great admirer of Bohr, Kramers was widely considered his faithful lieutenant, a position he willingly accepted.

At several occasions Kramers acted as Bohr's emissary when it came to disseminating and explaining the atomic theory to either physicists or the general public. For example, in 1917 and 1918 he went to Stockholm to expound the theory to Swedish scientists; in 1919 he was in Uppsala, and the following year in Lund. In Denmark he toured the country giving lectures to various groups of interested non-

*Natuurkunde*, 18 (1952), 161-166, reproduced in Aaserud 2007 (note 30), pp. 355-360. Neither in this address nor at any other occasion did Bohr mention the Kramers-Holst book.

specialists.[36] Although by 1922 he had written only a few popular works, he was already a seasoned populariser. What matters here is that in the early 1920s he was Bohr's closest collaborator and intimately familiar with Bohr's theory and his way of thinking. No one was better qualified than Kramers to write authoritatively about the Bohr atom.

Helge Holst is unknown internationally, but far from an uninteresting figure.[37] After having graduated in physics from Copenhagen University in 1893, followed by a brief period as assistant at the Polytechnic College, Holst turned to a career as writer and publisher of popular works in science and technology. The early years of the twentieth century was in Denmark (as in many other countries) a period in which there was a growing market for popular science. There was a great deal of interest among Danish scientists, educators and journalists to disseminate the marvels of modern science and technology to an attentive audience thirsting for knowledge.[38] This was a climate in which Holst, who was perhaps Denmark's leading and most productive populariser in the period, thrived. He was editor of the journal *Frem*, a popular journal of science and culture, and the author of a number of books, either alone or with coauthors. In 1920 Holst was appointed librarian at the Polytechnic College, in which position he remained until his death in 1944, continuing throughout his life to write articles and books on popular science and technology.

---

[36] Peter Robertson, *The Early Years: The Niels Bohr Institute 1921-1930* (Copenhagen: Akademisk Forlag, 1979), pp. 51, 95-97. Kramers to Bohr, 12 March 1917, in J. Rud Nielsen, ed., *Niels Bohr Collected Works*, vol. 3 (Amsterdam: Elsevier, 1976), p. 654. Kramers quickly learned to speak and write Danish.

[37] Hans M. Hansen, 'Helge Holst', *Fysisk Tidsskrift*, 43 (1945), 1-4 (in Danish).

[38] On popular science in Denmark in the early twentieth century, see Helge Kragh, Peter C. Kjærgaard, Henry Nielsen, and Kristian Hvidtfelt Nielsen, *Science in Denmark: A Thousand-Year History* (Aarhus: Aarhus University Press, 2008), pp. 357-383.

Holst's ambitions were not limited to disseminating science to the general public. He also had an interest in the foundation of physics, which caused him to revolt when he became acquainted with Einstein's new and controversial general theory of relativity. He was the first Dane to write comprehensively about Einstein's theory, which he did in 1919, characteristically in a literary and cultural magazine. The same year he published in German a lengthy memoir in the proceedings of the Danish Royal Academy of Sciences and Letters, which he followed up by a popular book in Danish and two papers in the *Zeitschrift für Physik*. In these works, which were primarily of a philosophical nature, he criticized Einstein's theory and advocated his own alternative of a 'causal relativity theory' based on the existence of a hypothetical 'neutral field' generated by the stars. Although Holst was not taken seriously by mainstream physicists, his views were known and discussed. 'I find the work by Helge Holst to be poor', Einstein wrote in a letter of 1920.[39]

In connection with his Royal Academy memoir of 1919, Holst was in contact with Bohr, whom he apparently (but in vain) tried to make interested in his views.[40] Nor would he have found any support from Kramers, who at the time was the only physicist in Denmark who had a solid technical knowledge of general relativity and in 1920 started giving regular courses on relativity theory to students at the University of Copenhagen.

---

[39] Einstein to Joseph Petzoldt, 21 July 1920, in Diana K. Buchwald et al., eds, *Collected Papers of Albert Einstein*, vol. 10 (Princeton: Princeton University Press, 2006), p. 341. In his authoritative review of relativity theory, Wolfgang Pauli referred to Holst's works, commenting that he found the hypothesis of a cosmic neutral field to be unnecessary. W. Pauli, *Relativitätstheorie* (Leipzig: Teubner, 1921), p. 560, a separate reprint of his article in *Encyclopädie der Mathematischen Wissenschaften*, vol. 5, part 2.

[40] Holst to Bohr, four letters of 1919, in Archive for History of Quantum Physics (AHQP), Bohr Scientific Correspondence (film 3, section 4). See also Holst to Kramers, 5 July 1921, AHQP, Kramers Correspondence M/f no. 8a Section 6-025. The information about Holst and the theory of relativity is in part based on an unpublished Master Thesis (Aarhus University) from 1998 by Jonas Cilieborg.

Finally, the Swedish physicist Oskar Klein, who replaced Kramers in writing and updating the Danish 1929 edition of *The Atom and the Bohr Theory*, was another of Bohr's close associates.[41] After having worked on and off with Bohr and Kramers from 1918 to 1923, he went to the University of Michigan, Ann Arbor, and in 1925 he returned to Copenhagen. There he did some very important work in theoretical physics, including a five-dimensional theory of quanta and relativity ('Kaluza-Klein theory' and 'Klein-Gordon equation'), an analysis of Paul Dirac's new theory of the electron ('Klein's paradox' and 'Klein-Nishina scattering') and a major contribution to the quantum theory of radiation ('Jordan-Klein theory'). In 1930 he left Denmark to become professor of physics at the Stockholm University College. In the present context it is relevant to point out not only that Klein was Bohr's faithful lieutenant (such as Kramers was), but also that he engaged in popularising the new atomic physics. For example, in 1922-1923 he wrote a comprehensive survey of Bohr's theory in *Kosmos*, a yearbook published by the Swedish Physics Association.[42]

## 5. The Atom and the Bohr Theory of its Structure

Before going into details about the content and impact of the book by Kramers and Holst (Figure 2), we first would like to mention its publication history. It was originally published in Danish in 1922, the same year that Bohr received the Nobel Prize.[43] However, at the time of writing, this was unknown, and so the book was not

---

[41] For Klein's life and career, see Abraham Pais, *The Genius of Science: A Portrait Gallery of Twentieth-Century Physicists* (Oxford: Oxford University Press, 2000), pp. 122-147.

[42] Oskar Klein, 'Den Bohrske Atomteorien', *Kosmos*, 2 (1922), 54-94 and 3 (1923), 72-120.

[43] The title was *Bohrs Atomteori Almenfatteligt Fremstillet*, meaning 'Bohr's Atomic Theory, a Popular Exposition'. The Danish term 'almenfattelig' corresponds to the German

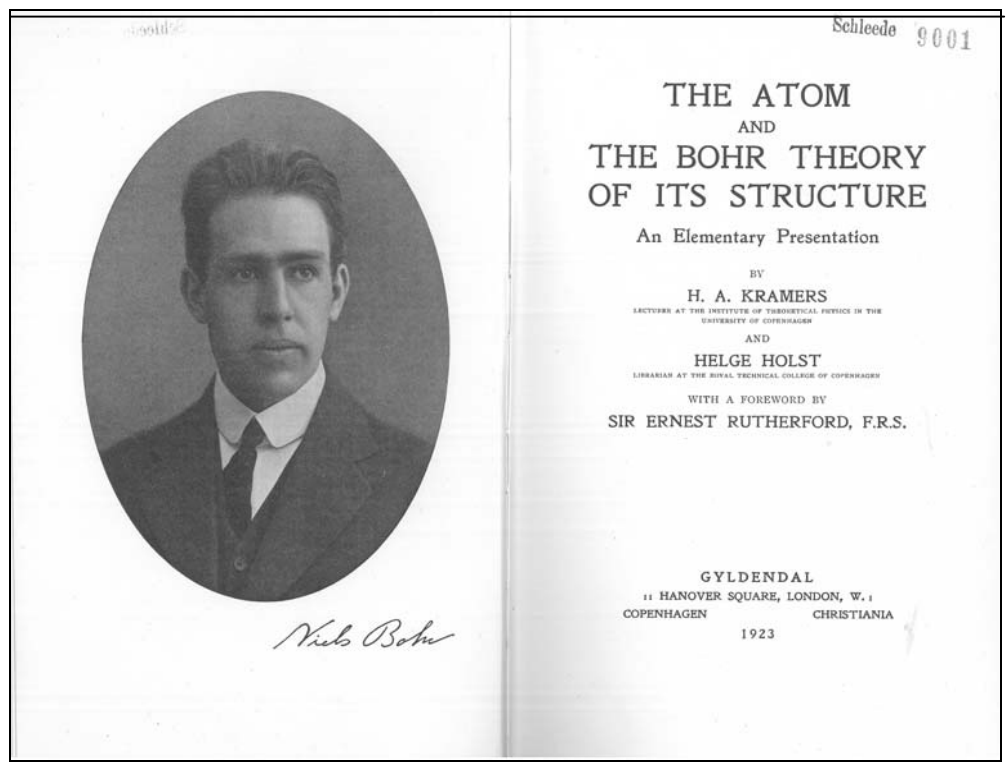

Fig. 2.  Frontispiece of Kramers and Holst's book of 1923, making no doubt
that atomic theory was the brainchild of Niels Bohr.

occasioned by the prestigious prize and the media attention it brought with it. The
publisher was the country's oldest and largest publishing house, Gyldendal, a
company named after its founder Søren Gyldendal who established it in 1770. There
is little doubt that the Nobel Prize was an important factor in the decision to make an
English translation, which in 1923 was published separately by Gyldendal in
England and by Alfred E. Knopf in the USA, in both cases with a foreword by
Rutherford. The translation was done by the US physics graduate Robert Bruce

'gemeinverständlich' or literally 'commonly intelligible'. In the Danish editions of 1922 and
1929 Holst appeared as a first author, while the order of the authors in the translations was
Kramers and Holst, presumably because Kramers was better known than Holst. The Danish
editions of 1922 and 1929 were both printed in 2000 copies. This should be seen in relation to
the population of the country, which in the 1920s was about 3.3 million, and may be
compared with the number of copies (3000) of Russell's *ABC of Atoms* (note 29).

Lindsay and his wife Rachel Tupper Lindsay, who spent the years 1922-1923 in Copenhagen on a stipend from the American-Scandinavian Foundation.[44] It may at a first blush seem strange that Bohr did not provide the Danish edition with a foreword, but this would hardly have been appropriate in a book that carried his name in the title and focused on his work. The lack of a foreword did not imply any lack of interest from Bohr's side.[45]

In 1925 *The Atom and the Bohr Theory* was translated into German and Spanish, and in 1927 it appeared in a Dutch translation, possibly occasioned by Kramers's return to the Netherlands the year before.[46] The Dutch version included sections on the new Heisenberg-Schrödinger quantum mechanics and differed in minor respects from the earlier versions.[47] The reason why the Kramers-Holst book was not immediately translated for the large German market may have been competition from similar books in German, in particular Kirchberger's *Entwicklung der Atomtheorie*.[48] In 1929 a second, updated edition appeared in Danish, now with the assistance of Klein. This edition left out parts of the 1922 edition in favour of two new sections: one on quantum mechanics and one on wave and particle descriptions of light and matter, both of which were primarily written by Klein. While the correspondence principle was given much attention in the first edition, in the second Bohr's new principle of complementarity was similarly highlighted.

---

[44] During his stay in Copenhagen, R. Bruce Lindsay worked under Bohr and Kramers, and the latter asked him to undertake the translation, much of which was actually done by Mrs. Lindsay. Apparently Kramers was himself involved in the translation. 'I have used the last two weeks on … the English translation of my book with Holst', he wrote to Bohr on 11 October 1923. See Rud Nielsen 1976 (note 36), p. 661. After his return to the USA, Lindsay completed his doctoral dissertation begun in Copenhagen and subsequently entered a distinguished career in physics as professor at Brown University.

[45] Bohr did write a foreword to a later popular book on atomic physics coming from his institute and written by two of his collaborators, Christian Møller and Ebbe Rasmussen. This book, *Atomer og Andre Småting* [Atoms and Other Small Things] (Copenhagen: Hisrchsprung, 1938) followed the general structure of the Kramers-Holst book, but of course extended with aspects of quantum mechanics, nuclear physics and other post-1925 developments.

[46] The German edition: *Das Atom und die Bohrsche Theorie seines Baues* (Berlin: Springer, 1925), translated by Fritz Arndt, a professor of chemistry at the University of Breslau. The Spanish edition: *El Átomo y su Estructura Según la Teoria de N. Bohr* (Madrid: Revista de Occidente, 1925), translated by Tomás R. Bachiller. The Dutch edition: *De Bouw der Atomen* (Amsterdam: D. B. Centen, 1927), translated by Henri C. Brinkman, reprinted 1930 and in 1949 under the title *De Bouw der Atomen en Moleculen*.

[47] Kramers contemplated writing a Dutch translation as early as 1923, which appears from Holst to Kramers, 29 July 1923, AHQP, M/f No. 8a, Sect. 6-026.

The prefaces of the two publications of 1922 and 1923 presented Bohr slightly differently. The Danish edition referred to Bohr as 'the young Dane', who 'in 1913 advanced a theory that not only provided a surprisingly simple explanation of certain physical facts, in the face of which physics until then had remained perplexed, but also offered undreamt-of possibilities for future research'. According to the British edition, on the other hand, Bohr simply 'paved the way for a really physical investigation of the problem', namely, the problem of explaining in terms of general laws the physical and chemical properties of the elements. The Danish preface furthermore stated that Bohr, because of the 'revolutionary character' of his ideas, 'for a while had to walk alone, guided by his extensive knowledge, his great power of combination and, not in the least, his certain instinct'. Joined by other physicists in his quest for the secrets of the atom, Bohr for a while almost seemed as if he was lacking behind the others, but, truly, the Danish preface boldly declared, 'Bohr was searching in the dark far ahead', and thus, 'physical research today is carried out under the sign of the Bohrian theory'. Attributing slightly less heroism to Bohr, the English preface fully acknowledged the contribution of Niels Bohr:

---

[48] This is what Walther Grotrian suggested in a review of 1925: W. Grotrian, *Die Naturwissenschaften*, 13 (1925), 952-953.

The past decade has witnessed an enormous development at the hands of scientists in all parts of the world of Bohr's original conceptions; but through it all Bohr has remained the leading spirit, and the theory which, at the present time, gives the most comprehensive view of atomic structure may, therefore, most properly bear the name of Bohr.

The book consists of seven chapters, the first four of which detail important theories and discoveries in the history of physics and chemistry leading up to Bohr's theory of the atom. There is mention of Dalton's atomic theory, Mendeleev's development of the period table of elements, Maxwell's electromagnetism, Balmer's and Ritz's formulae of atomic spectra (the book contained colour plates of spectra produced by Bunsen and Kirchhoff[49]), and Thomson's and Rutherford's discoveries of the electron and the nucleus, respectively. The two authors did not fail to point out several contributions to physics made by Danish scientists, such as Ludvig A. Colding's work of the 1840s anticipating the principle of energy conversation, Christian Christiansen's experiments with blackbody radiation from the early 1880s, and Martin Knudsen's work from about 1915 on gases at very low pressure.[50]

In their outline of the core of Bohr's theory, Kramers and Holst dwelled upon the two postulates or what they called 'fundamental concepts', that is, the constraint of stationary states and the frequency condition.[51] They depicted the Bohr model of the hydrogen atom in a simplified form, a point-like nucleus surrounded by electrons moving circularly (or elliptically) in stationary orbits. In each of these orbits, they explained, the electron follows the general mechanical laws of motion, but it contradicts classical electrodynamics by emitting no electromagnetic waves: radiation is only emitted when the electron passes from one orbit to another.

---

[49]  These plates of spectra, going back to the early 1860s, were reprinted in numerous texts on spectrum analysis. According to Klaus Hentschel, 'it was the most frequently reprinted scientific illustration in the second half of the nineteenth century'. K. Hentschel, *Mapping the Spectrum: Techniques of Visual Representation in Research and Teaching* (Oxford: Oxford University Press, 2002), p. 48.

[50] Kramers and Holst (note 9), p. 25 and p. 33. Christiansen was Bohr's teacher and served as professor of physics at Copenhagen University 1886-1912, after which he was followed by Knudsen.

From the beginning of their outline of Bohr's theory, Kramers and Holst emphasised Bohr's attempts to preserve and develop the connection between quantum theory and classical physical theories and observations. They described the correspondence principle as 'difficult to explain …, because it cannot be expressed in exact quantitative laws, and it is, on this account, also difficult to apply'.[52] Nevertheless, the authors cited several applications of the correspondence principle, according to which there is a formal correspondence between the states described by elliptic electron orbits and the radiation emitted (or absorbed) due to transitions between the states. However, they also pointed out that it is not obvious why the motion of an electron in a stationary state should be described in classical terms when the transition between states is thoroughly unclassical. Kramers and Holst, closely following Bohr's own interpretation, put it this way:

We then also here see the outward similarity between the Bohr theory and the classical electrodynamics. We may say that the radiation of frequency $\nu$, produced by a single jump, *corresponds* to the fundamental harmonic component in the motion of the electron, while the radiation of frequency $\nu_2$, emitted by a double jump, corresponds to the first overtone, etc. The similarity is, however, only of a formal nature, since the processes of radiation, according to the Bohr theory, are of a quite different nature than would be expected from the laws of electrodynamics.[53]

---

[51] Ibid., pp. 117-118, 138.
[52] Ibid., p. 139.

Unsurprisingly, given Kramers's involvement in the formulation and application of the principle of correspondence, the two authors described the principle in considerable detail, presenting it as 'one of Bohr's deepest thoughts and chief guides'. The principle of correspondence between quantum and classical theory not only had turned out be 'extraordinarily fruitful' for atomic physicists, but also 'has made possible a more consistent presentation of the whole theory, and it bids fair to remain the keystone of its future development'.[54]

Like most other authors describing elements of the new atomic theory to a general audience, Kramers and Holst used a classical analogy to illustrate the Bohr atom. They asked their readers to compare the atom with a hypothetical musical instrument consisting of a series of circular discs placed one over another, each disc being smaller than the one above.[55] A ball would move frictionless around any of the discs, corresponding to a system in a stationary state. The ball might fall down to any lower disc, emitting a sound. Passing from one stationary state to another, the system would lose a quantity of energy equal to the work necessary to raise the ball again. The energy lost by moving from one state to another would be emitted as a sound from the instrument. If the smallest disc was grooved in such a way that the ball could fall no further, then 'this fanciful instrument can provide a rough analogy with the Bohr atom. We must beware, however, of stretching the analogy farther than is here indicated.'[56] The analogy nicely illustrated the correspondence between an atomic system and classical descriptions, but it also indicated some of the problems of applying the correspondence principle. As Kramers and Holst pointed out, correspondence considerations are applicable only to certain aspects of the theory, not all of them. Moreover, the theory itself says nothing about when to apply the principle of correspondence. It was to some extent an ad hoc solution to a host of theoretical problems, and, furthermore, it raised a number of additional problems.

---

[53] Ibid., pp. 130-131. Other aspects of correspondence between quantum and classical theory mentioned in the book included the formal agreement between the Balmer-Ritz formula for the hydrogen spectrum and Bohr's quantization postulate, the agreement between calculations of the magnitude of ionisation potentials by means of quantum and classical theory, the quantum theory-based derivation of the Rydberg constant, and the calculation of 'more complicated electron motions than those which appears in the unperturbed hydrogen atom' (p. 141).

[54] Ibid., p. 141.

[55] The acoustical analogy was probably due to Holst, who in a paper from the spring of 1922 introduced it to illustrate Bohr's idea of emission of radiation. H. Holst, 'Om Niels Bohrs Værk', *Tilskueren* (May 1922), 281-287.

Always eager to spot theoretical problems, Niels Bohr was keen to discuss difficult issues in relation to quantum theory and the correspondence principle. Oskar Klein recalled that, in conversations, Bohr 'preferred to take up unsolved problems, around which his thought moved incessantly'.[57] So, most likely, Kramers and Holst were inspired by Bohr's fondness of open questions when they chose to discuss particular problems of Bohr's theory in their book. Also, the way in which they framed their discussion of the new difficulties was typical of Bohr's thinking. They never, as did, for example, Bertrand Russell in his *ABC of Atoms*, said that the problems would probably be solved by physicists in the future, nor did they, as was also common in popular books about the atom, simply neglect problematic issues. Rather, in line with Bohr's own approach, they made a point of warning against the idea that the Bohr theory could be used to derive 'everything that happens in the atom and so all in nature'.[58]

---

[56] Kramers and Holst 1923 (note 9), p. 120.
[57] O. Klein, 'Glimpses of Niels Bohr as Scientist and Thinker', pp. 74-93 in Stefan Rozental, ed., *Niels Bohr: His Life and Work as Seen by his Friends and Colleagues* (Amsterdam: North-Holland, 1967), on p. 77.

Ever since 1913, critics of Bohr's theory had objected to what they considered the serious conceptual and methodological problems associated with the theory.[59] Kramers and Holst did not eschew these problems, although neither did they conclude, as some critics did, that they were reasons to disbelieve Bohr's view of the constitution of atoms. As Bohr had himself emphasised, his theory offered no *explanation* in the ordinary sense of either the stationary states or the jumps between them. Echoing their master, Kramers and Holst wrote: 'We are inconceivably far from being able to give a description of the atomic mechanism, such as would enable us to follow, for example, an electron from place to place during its entire motion, or to consider the stationary states as links in the whole instead of isolated "gifts from above".'[60]

Among the problems dealt with by Kramers and Holst was the peculiar fact that the electron, in making a transition between two, non-adjacent energy levels, will not emit all the intervening frequencies, but simply the frequency corresponding to the entire jump. Right from the beginning of the jump, 'the electron seems to arrange its conduct according to the goal of its motion and also according to future events. But such a gift is wont to be the privilege of thinking beings that can anticipate certain future events.'[61] In other words, the electron might seem to be endowed with a kind of free will, its motion being determined teleologically rather than causally. This was an old problem, first pointed out by Rutherford in a letter to Bohr of March 1913,[62] and Kramers and Holst made no attempt to solve it. Following Bohr, they chose to consider the paradoxical behaviour of atomic electrons a stimulating challenge to our thinking about the subatomic world rather than a real problem for the Bohr atom. It indicated that it might be 'impossible to obtain a consistent picture of atomic processes in space and time.'

---

[58] Kramers and Holst 1923 (note 9), p. 132.
[59] On these problems, see Helge Kragh, 'Conceptual Objections to the Bohr Atomic Theory – Do Electrons Have a Free Will?', *European Physical Journal H* (forthcoming).
[60] Kramers and Holst 1923 (note 9), p. 133. According to Bohr, his atomic theory 'does not attempt an "explanation" in the usual sense of this word, but only the establishment of a connexion between facts which in the present state of science are unexplained'. Bohr 1922 (note 31), p. v.
[61] Kramers and Holst 1923 (note 9), p. 136.

By presenting the reader with an analogy of the atom and then proceed to discuss its weaknesses, Kramers and Holst illustrated Bohr's own way of working. Bohr liked to fully explore the logical consequences and contradictions of any model or concept, only in order to come up with new models and new concepts, which could also be explored in terms of their weaknesses and ambiguities. For Bohr, this was simply the way he worked. In a popular book about science, however, the discussion of uncertainties and unknowns in the theory of atoms had to be limited. Most likely because of their lengthy presentation of some of the shortcomings of the Bohr model, Kramers and Holst had to warn their readers against 'the impression that the Bohr theory, while it gives us a glimpse into depths previously unsuspected, at the same time leads us into a fog, where it is impossible to find the way'.[63] On the contrary, they argued, the best proof that Bohr's theory was no blind alley for physicists was its ability to predict and account for many phenomena with remarkable accuracy and in complete agreement with experimental observations. To make their point, the two authors, having presented the Bohr theory, turned their attention to 'the first great triumphs in which the theory showed its ability to lead the way where previously there had been no path'.[64] These triumphs included the spectroscopic verifications of the theory in the case of the hydrogen atom and also the construction of atomic models of the higher atoms in agreement with the periodic system. Before turning to the atomic models and their pictorial representations, we want to comment on a section in *Das Atom und die Bohrsche Theorie* which did not appear in the earlier English edition.

---

[62] Rutherford to Bohr, 20 March 1913, in Ulrich Hoyer, ed., *Niels Bohr. Collected Works*, vol. 2 (Amsterdam: North-Holland, 1981), p. 583. For more on this problem, see Kragh (note 59).
[63] Ibid., p. 138.

Due to the rapid progress of quantum theory, by 1925 the Bohr atom was not quite the same as it had been two years earlier. In an attempt to update the content of the German edition, it was expanded with a new chapter on the interaction of light and matter written by Kramers. This chapter is of interest because it gives a clear and non-technical account of how Bohr and his assistants in Copenhagen looked at the radiation problem, which in the last phase of the old quantum theory became increasingly important. To make a long story short, by late 1923 Einstein's idea of localized light quanta (later known as photons) had become accepted by a substantial part of the physics community, but not by Bohr and Kramers in Copenhagen. Based on an idea of 'virtual oscillators' proposed by the US physicist John Slater, who stayed at Bohr's institute from December 1923 to June 1924, Bohr and Kramers proposed an alternative theory of light emission based on the wave picture of light and the radical hypothesis that energy and momentum are conserved only statistically. With energy nonconservation followed the equally radical idea that the principle of causality might not be valid in the subatomic domain.

The short-lived BKS (Bohr-Kramers-Slater) theory aroused great attention in the physics community, if little sympathy outside Copenhagen, and in May 1925 it was disproved by experiments made by Walther Bothe and Hans Geiger in Berlin.[65] When Kramers wrote his chapter to the German edition, the preface of which was dated March 1925, the theory was still alive and it figured prominently in his exposition.

---

[64] Ibid., p. 142.

While certainly hoping that the BKS theory (or what he consistently called 'Bohr's new view') would turn out to be correct, Kramers emphasised its unfinished nature and modestly characterised it as 'only an attempt to throw a little light in the great darkness of our ignorance about the course of the atomic processes … essentially a working-programme for the theorists'. As to the element of acausality, he preferred to consider it 'rather a matter of taste', although there is little doubt that his own taste (and Bohr's as well) was acausal. To Kramers, the principle of causality was a fact of experience rather than a logical necessity, and 'one could easily imagine that it breaks down for atomic processes'. Similarly, one should keep an open mind with respect to a violation of the law of energy conservation. Interestingly, Kramers suggested that large-scale energy nonconservation processes might go on here and now. There are indications, he said, that 'in the hot stars, … the principle of energy conservation cannot be used just like that, but that in these bodies there occurs, so to speak, a spontaneous creation of energy which contributes to maintain the enormous radiation of energy that the stars pour into space'.[66]

---

[65] Much has been written about the BKS theory and its role in the final phase of the old quantum theory. See, for example, Dresden (note 8), pp. 41-78, 159-215 and Sandro Petruccioli, *Atoms, Metaphors and Paradoxes: Niels Bohr and the Construction of a New Physics* (Cambridge: Cambridge University Press, 1993), pp. 111-133. According to Dresden, the treatment of the BKS theory in the Kramers-Holst book 'is without much doubt the most understandable exposition of the BKS ideas' (p. 195).

[66] Kramers and Holst 1925 (German edition, note 46), pp. 123-140. The chapter also appeared as a separate article in Danish, see H. A. Kramers, 'Om Vekselvirkningen mellem Stof og Lys', *Fysisk Tidsskrift*, 23 (1925), 26-40. The idea that stellar energy is rooted in processes violating energy conservation was later taken up by Bohr, who advocated it from about 1929

## 6. Pictorial atoms

In his Nobel lecture of 1922 as well as at other occasions Bohr had accounted for the main features of the periodic system of the elements by means of atomic models based on two quantum numbers. While in 1913 he only made use of the 'principal' quantum number $n$ (with integral values 1, 2, 3, …), to describe elliptic orbits it was necessary to take into account also the 'azimuthal' quantum number $k$ introduced by Sommerfeld in his 1915 generalization of Bohr's theory. An electronic orbit could then be characterized as an $n_k$ orbit, where $k$ can attain the values 1, 2, …, $n$. Geometrically, the greater the difference between the two quantum numbers, the greater the eccentricity of the elliptic orbit. Only in the case $k = n$ is the eccentricity zero, meaning that the ellipse degenerates into a circle.

Kramers and Holst described Bohr's two-quantum construction of atoms in some detail, providing it with a number of pictures placed at the end of the book (Figure 3). Orbits with even $n$ were drawn in black, those with odd $n$ in red, and all the orbits were roughly drawn to scale. Kramers and Holst noted that their diagrams, seductively looking like images of real atoms, should not be taken too literally: 'Although the attempt has been made to give a true picture of these orbits as regards their dimensions, the drawings must still be considered as largely symbolic. Thus in reality the orbits do not lie in the same plane, but are oriented in different ways in space.' In line with their previous discussions of the shortcomings of the quantum

---

to 1933. Well aware that the idea was speculative, Kramers did not expound it in any of his scientific publications. Indeed, one of the functions of popular works is that scientists have greater liberty in suggesting ideas of a speculative nature that would not be found acceptable in scientific articles.

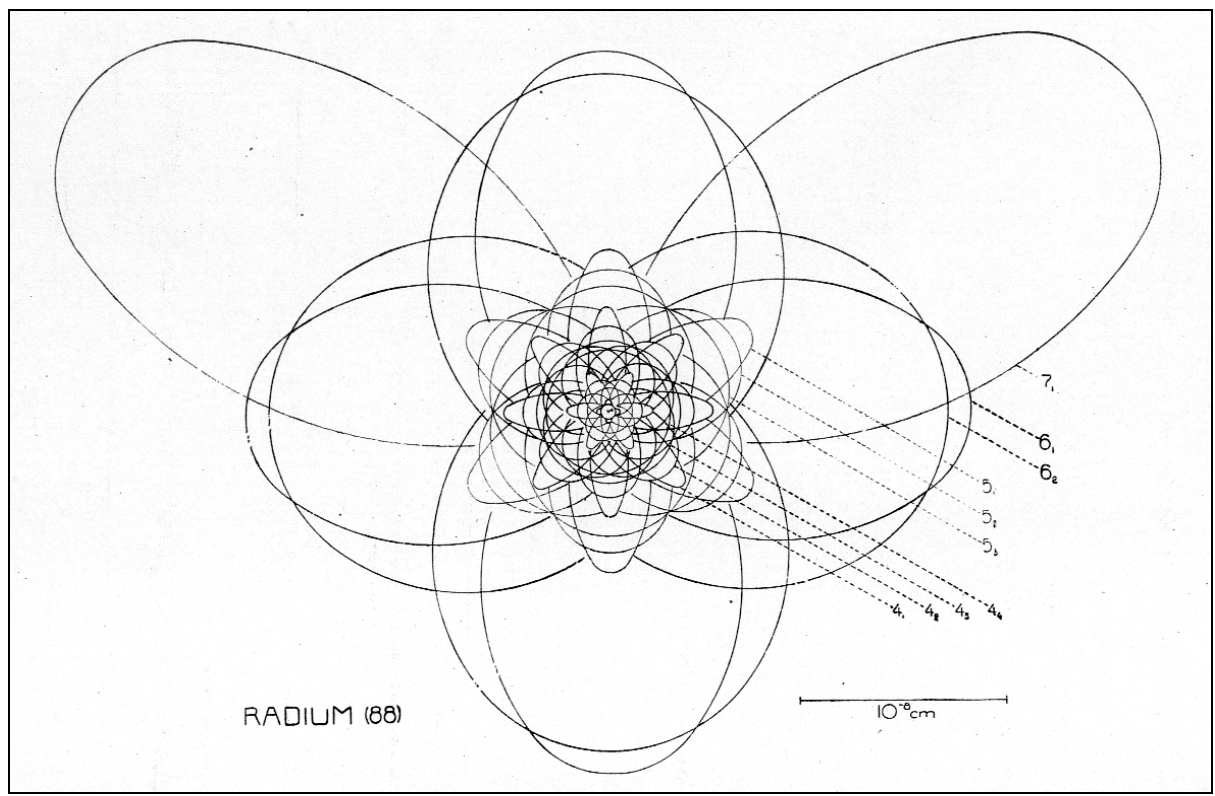

Fig. 3. The 88 electron orbits of a radium atom according to Bohr's theory of 1921. The elliptic orbits are shown closed for simplicity, but should really be slightly open, as the ellipses slowly precess. Source: Kramers and Holst 1923 (note 9).

theory, they further noted that there was 'still a good deal of uncertainty as to the relative positions of these planes'.[67]

　　Little is known about the origin of the diagrams, but apparently they were made on Bohr's request for his lectures and not specifically for the book.[68] The table with the pictures of atoms was included in both the Danish, English, German, and Spanish editions (but not in the Dutch edition of 1927). However, while the Danish edition of 1922 included eleven elements (H, He, Li, C, Ne, Na, Ar, Kr, Cu, Xe, and Ra), the carbon atom with its 'beautiful spatial symmetry' of four $2_1$ valence electrons was missing in the other editions. Why? We suspect that the reason was that Bohr had come to doubt his symmetric configuration of the carbon atom, which soon turned out to be two $2_1$ electrons and two $2_2$ electrons in the outer shell.[69] At any rate, Kramers and Holst took an interest in the reproduction of the diagrams in the translated editions. Thus, in a letter to Kramers of 29 July 1923, Holst expressed concern that the format of the British, German and Dutch translations would not be big enough to allow the diagrams to be included. If not, new blocks probably would have to be made.[70]

[67] Kramers and Holst 1923 (note 9), p. 192.
[68] This information comes from H. A. Kramers, 'Das Korrespondenzprinzip und der Schalenbau des Atoms', *Naturwissenschaften*, 11 (1923), 550-559, on p. 556. The Niels Bohr Archive in Copenhagen possesses a number of glass slides from the period with atomic pictures, but it is unknown whether these are the originals used by Bohr or reproductions from the Kramers-Holst book.

The coloured plates with pictures of atoms that were attached to the book by Kramers and Holst undoubtedly appealed to many readers and were also eminently useful for public presentations. They were reused at a number of occasions, first by Kramers in an article in a special issue of *Naturwissenschaften*, published in July 1923 and celebrating the first decade of Bohr's theory. Apart from advertising the popular Danish book, Kramers repeated that the pictures should not be regarded as true representations of atoms. They were merely meant as 'a rough illustration'.[71] Shortly later the Canadian physicist John McLennan, at the University of Toronto, used the pictures in his address to the Liverpool meeting of the British Association in September 1923 in which he gave a careful presentation of Bohr's recent ideas of atomic structure. The pictures 'have been copied from a paper by Kramers that has recently appeared and are stated to be similar to those prepared by Bohr for use in his own lectures'.[72] The English translation of the Kramers-Holst book had not yet appeared.

[69] Bohr admitted his mistake in an appendix to the second (1924) edition of Bohr 1922 (note 31), see p. 138. Whereas the symmetric structure of the carbon atom was carefully described in the 1922 edition of the Kramers-Holst book, it was not mentioned in the later editions.
[70] Holst to Kramers, 29 July 1923, AHQP, M/f No. 8a, Sect. 6-026. See also Holst to Kramers, 7 May 1925, AHQP, M/f No. 8a, Sect. 6-027, concerning the plates in the Spanish edition.
[71] Kramers 1923 (note 68), p. 556. Reference to the plates and the Kramers-Holst book was also made by the Dutch physicist Dirk Coster in his contribution to the special issue. Characteristically, Coster worked at the time at Bohr's institute.

The Kramers-Holst pictorial models of atoms can be found in several other cases of the popular or semi-popular literature, not always with reference and rarely with permission. Andrade were unaware of the plates when he wrote his *Structure of the Atom*, but the following year, in a contribution to a general work celebrating the progress in chemistry, he included pictures of some of the simpler atomic models (He, Ar, Ne, Na).[73] The atomic models also appeared in a popular book written in 1924 by Lars Vegard, professor of physics at the University of Oslo (then Kristiania), who was internationally known for his research on the physics of the aurora borealis.[74]

The same year the atomic pictures turned up in an article in the Spanish popular science journal *Iberíca* and probably many other places. The article in *Iberíca* was a translation of Rutherford's opening address to the 1923 British Association meeting, which did not in fact include any pictures or reference to them. The editors of the Spanish journal pasted the pictures of the atoms to the article to make it more inviting.[75]

---

[72] John C. McLennan, 'On the Origin of Spectra', *Report, British Association of the Advancement of science* (London: J. Murray, 1924), 25-58.

[73] E. Andrade, 'The Structure of the Atom', pp. 43-55 in Edward F. Armstrong, ed., *Chemistry in the Twentieth Century* (London: Ernest Benn, 1924), on p. 53.

[74] Lars Vegard, *Stoffets Opbygning og Atomenes Indre* (Kristiania: Olaf Norlis Forlag, 1924). Although Vegard paid tribute to "the Danish physics-genius Niels Bohr," his book was less Bohr-focused than most popular physics books at the time.

[75] The Spanish translation of Rutherford's address appeared in four sequels in *Iberíca* nos. 357, 542, 545, and 551. On this journal, see Maria C. Boscá, 'Some Notes on the Popularization of Quantum and Atomic Physics in Spain, 1914-1927', pp. 61-74 in Arne

As a last example, the Kramers-Holst pictures turned up in the printed version in the *Bayerische Radio-Zeitung* of a lecture broadcasted by the Polish-German radiochemist Kasimir Fajans.[76] It should be mentioned that the Copenhageners were not alone in presenting visual models of the atom for pedagogical and educational purposes. Thus, as early as 1918 Sommerfeld made a sketch of the hydrogen atom which a few years later, together with a model of the positive hydrogen molecule ion, was turned into a three-dimensional model displayed at the Deutsches Museum in Munich.

Similar but more complicated models of the Bohr orbital atom, based on calculations made by Lawrence Bragg and Douglas Hartree, were shown at the British Empire Exhibition in London 1924-1925 and later at the Science Museum.[77] Moreover, the Dutch physicist Heike Kamerlingh Onnes, a Nobel laureate of 1913 for his fundamental research in low-temperature physics, made use of pictorial atoms in his attempt to understand why superconductivity is restricted to a few metals (Figure 4). He received the pictures, which were more schematic versions of the Kramers-Holst models, from Kramers in Copenhagen.[78]

---

Schirrmacher, ed., *Communicating Science in the 20th Century*. Max Planck Institute for the History of Science, Preprint 385.

[76] We have this information from an unpublished talk of 2007 by Arne Schirrmacher, see slide 22 of the power point presentation included in http://quantum-history.mpiwg-berlin.mpg.de/eLibrary/hq1_talks/old-qt/06_schirrmacher.

[77] For the models in Munich and London, see Schirrmacher 2009 (note 8).

[78] Heike Kamerlingh Onnes, 'Rapports sur de nouvelles expériences avec les supraconducteurs', *Communications from the Physical Laboratory of the University of Leiden*, 1924, Supplement 50a. For Kamerlingh Onnes's interest in atomic models, see Tilman Sauer, 'Einstein and the Early Theory of Superconductivity', *Archive for History of Exact Sciences*, 61 (2007), 159-211.

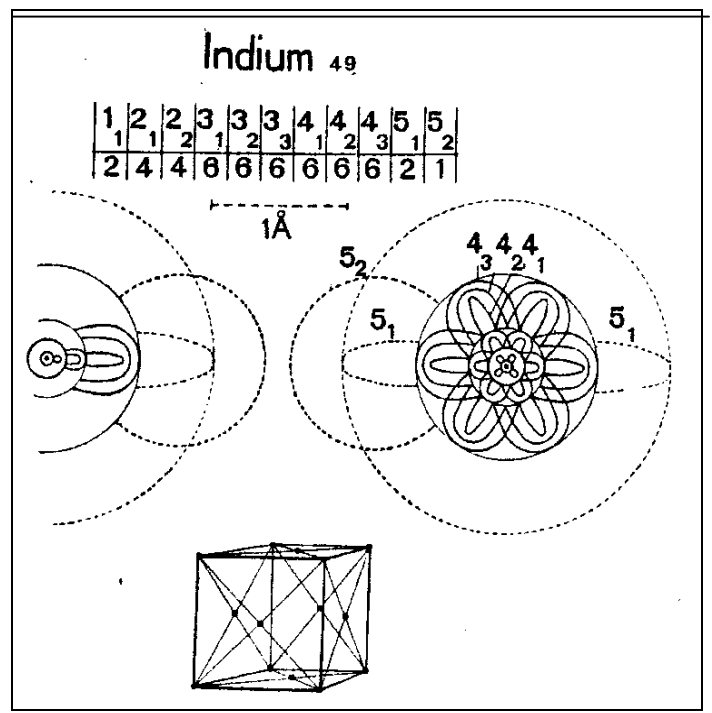

Fig. 4. The electronic and lattice structure of indium, according to Bohr's atomic theory and as used by Kamerlingh Onnes in a paper on superconducting metals. Source: Kamerlingh Onnes 1924 (note 78).

## 7. Reception and dissemination

The Kramers-Holst volume was successful and well received in the physics community as a fine example of popular science. According to Rutherford's foreword to the English translation, the book was not only 'a clearly written and accurate account' of Bohr's atomic theory, it was also commendable because of its plain language and lack of mathematics. 'This book', Rutherford said, 'should prove attractive not only to the general scientific reader, but also to the student who wishes to gain a broad general idea of this subject before entering into the details of the mathematical theory'.[79] As noted by an anonymous reviewer in *Nature*, the book differed from other popular works on atomic theory by its focus on and devotion to Bohr's way of conceiving quantum and atomic physics: 'We have a discussion in every part of which the spirit of the Bohr theory walks abroad.' The reviewer agreed with Rutherford as to the book's pedagogical qualities and also that its use was not restricted to a general readership. The quantum atom was widely seen as abstruse, but 'the book [is] extremely valuable to the not inconsiderable number of physicists who feel the need of a general and authoritative account of the latest speculations on these matters'.[80]

---

[79] Foreword dated 8 October 1923.

Some of the themes mentioned in the *Nature* review essay also appeared in a longer and highly appreciative review by George L. Clark, a chemist at the Massachusetts Institute of Technology and a specialist in X-ray analysis. 'It is almost as if … Bohr himself were speaking – as no doubt he is', Clark commented, suggesting that 'the two Danes, Kramers and Holst' engaged in a kind of teleological historiography by presenting the history of atomism as setting the stage for 'the advent of the immeasurably great contribution of Niels Bohr'. Like some other scientists reviewing the book, Clark thought (probably unrealistically) that it was a work accessible to the proverbial man on the street:

> It can be understood by anyone with average intelligence; in fact, it should prove fascinating to all ages from twelve to fourteen up. … It is a book for the home library table, for the college lecture room, for the scientist's reference shelf, for the minister who would preach truth and faith. It is the kind of book which may be read appreciatively at a single sitting in two or three hours, or in bits, even on a street car (the reviewer has observed this in three instances).[81]

---

[80] Review essay, 'Science for the People', *Nature*, 113 (1924), 378-380, of Kramers and Holst 1923 (note 9) and Sullivan (note 27). The assessment that the book, although of a popular nature, was valuable to scientists because of its authoritative statement of Bohr's view, was repeated in a review by Frank Hoyt, a US physicist who knew Kramers from his stay in Copenhagen 1922-1924: *Astrophysical Journal*, 61 (1925), 453.

Bohr spent the late fall of 1923 in the USA and Canada, lecturing on his atomic theory at several universities. The prestigious Silliman Lectures that he gave at Yale University between 6 and 15 November attracted much attention and were covered by the *The New York Times*. The day after the first lecture the newspaper informed its readers that 'Likened to solar system, he [Bohr] pictures the atom with nucleus corresponding to sun, and electrons to planets'. On 27 January 1924 it noted in its book section that 'The recent lectures at Yale and Columbia by the Danish scientist, Niels Bohr, … lend additional interest to the publication … of "The Atom and the Bohr Theory of Its Structure" by Helge Holst and H. A. Kramers'.

Given that the Kramers-Holst book was practically devoid of mathematics, it is a bit curious that it was reviewed in at least two mathematical journals. One of these reviews, by the Viennese physicist Otto Halpern, found it to be 'an easy and understandable read' but also pointed out that Kramers's chapter on radiation theory failed to mention the Compton effect, which he considered 'a major deficiency'.[82] Indeed it was, for the scattering effect discovered by Arthur Compton in 1923 was generally taken as strong (if not compelling) evidence for the light quantum. But then neither Kramers nor Bohr believed in the light quantum and they also did not believe that the Compton effect proved its existence since the BKS theory provided an alternative explanation. Still, to write about light quanta and the nature of light in the spring of 1925 without mentioning the Compton effect was an aberration that might smell of partisanship.

---

[81] *Journal of the American Chemical Society*, 46 (1924), 1318-1319.
[82] Otto Halpern, *Monatshefte für Mathematik*, 35 (1925), 32-33. The other review, of the English edition, was by R. D. Carmichael and appeared in *Bulletin of the American Mathematical Society 30* (1924), 374.

We finally mention a review by Walter Grotrian, a Berlin astrophysicist and specialist in spectroscopy, who mistakenly believed that both authors 'belong to the closely knit circle around Niels Bohr'. Kramers was 'the well-known collaborator of N. Bohr' and Holst was thought to belong to the same group. Comparing the book to Kirchberger's, Grotrian suggested that the Kramers-Holst work was a better buy because it combined a truly popular (*gemeinverständlich*) exposition with an authoritative account of Bohr's atomic theory.[83] In stark contrast to Einstein's theory of relativity, Bohr's theory of the atom received almost no attention from philosophers. Russell was an exception, but then his *ABC of Atoms* was not philosophical in nature. Another of the few exceptions was Harold Chapman Brown, a philosopher at Stanford University. Being 'fearful of getting beyond my depths in the intricacies of modern physics', he took recourse to *The Atom and the Bohr Theory of its Structure* from which he cited a passage indicating the complete lack of knowledge about an atom during a transition from one stationary state to another. Brown found this to be a revolutionary speculation, for 'it deprives matter of its eternal existence, an essential attribute under the old conception'.[84]

While the general public, when presented with the Kramers-Holst pictures, could hardly avoid believing that these were nearly authentic representations of what atoms really look like, specialists in atomic theory were well aware that a model should not be confused with reality. Although Bohr and Kramers considered the pictures as symbolic rather than concrete representations, still in 1923 they had little doubt about the reality of the electron orbits. Sure, the atom did not look like the picture, but it might still be something like it. Other physicists, and especially the youngsters Pauli and Heisenberg, held more radical views, doubting the very legitimacy of electron orbits.

---

[83] Grotrian (note 48). It should be mentioned that Dresden (note 8), p. 134, refers to a 1924 review by Max von Laue, also in *Naturwissenschaften*. This is puzzling, for no such review exists and yet Dresden quotes from it!

[84] Harold C. Brown, 'The Material World – Snark or Boojum?' *Journal of Philosophy*, 22 (1925), 197-214 (on p. 203), based on an address delivered to the American Philosophical Association on 28 November 1924. The revolutionary passage that aroused the philosopher's attention appeared in Bohr and Kramers (note 9), pp. 133-134.

By the summer of 1924 the visual analogue of the Bohr or Bohr-Sommerfeld atomic model was fading and no longer considered as a viable image of the real structure of atoms. The constitution of the atom in terms of a tiny positive nucleus surrounded at great distances by a system of electrons was left untouched, and so was the postulate of stationary states; but few leading physicists believed in the planetary analogy, that the electrons actually moved in definite orbits whose geometry was characterized by quantum numbers. Objections to the orbital model had been around for some time, raised in particular by Pauli, who reached the conclusion that 'the conception of definite and unambiguously determined electron orbits in the atom can hardly be sustained', as he wrote to Sommerfeld.[85] It was as if such models emotionally offended him, and he seems to have associated them with Kramers rather than Bohr. In a letter to Bohr of 12 December 1924 he poked fun at 'our good friend Kramers and his colourful picture books', obviously a reference to the Kramers-Holst book and its pictorial atomic models.[86] As Pauli saw it, in so far one could speak of atomic models at all, it had to be a mathematical and not a pictorial model.

[85] Pauli to Sommerfeld, 6 December 1924, in Michael Eckert and Karl Märker, eds, *Arnold Sommerfeld. Wissenschaftlicher Briefwechsel*, vol. 2 (Berlin: Verlag für Geschichte der Naturwissenschaften und der Technik, 2004), p. 177.

[86] Pauli to Bohr, 12 December 1924, in Klaus Stolzenburg, ed., *Niels Bohr. Collected Works*, vol. 5 (Amsterdam: North-Holland, 1984), p. 427.

Of course, with the advent of quantum mechanics Pauli's view became generally accepted and the pictorial orbital models lost whatever credibility there was left. Yet the pictures lived on in the Danish 1929 edition of the Kramers-Holst book which was revised by Oskar Klein and supplemented with a chapter on the new quantum mechanics and Bohr's interpretation of it. The two-colour plates, scientifically unjustified as they were, were apparently found to be too good to be left out.

### 8. Conclusion: The Kramers-Holst book in a 'popular science' context

While we begin to have a historical understanding of the scientific reception of the Bohr theory in different European contexts, the popular exposition and reception of the Bohr theory still remains relatively unexplored. Moreover, we have little understanding of the interaction between atomic physics proper and popular physics in this period. It is relatively easy to understand why some physicists and science popularisers attempted to make quantum theory comprehensible in non-mathematical language and visual images. They viewed popular science as a way in which to propound some of the latest scientific ideas about the structure of the atom; physicists did so as they were grappling with the mathematical and conceptual development of quantum theory, science popularisers as they were struggling to meet (and spark) the (perceived) popular demand for new scientific knowledge.

However, we currently lack understanding when it comes to what non-scientific audiences made of popular representations of quantum theory, and when it comes to the contribution of popular science to the social and cognitive dynamics of the scientific community. Case studies from the later part of twentieth century have shown that popular science can be a useful tool for scientists who want to attract attention to emergent scientific ideas by distributing them across the disciplines as widely as possible.[87] We acknowledge that physicists, earlier on, might have had similar strategic interests in popular science, such as spreading the gospel of Bohr's theory, but we also want to maintain a more heterogeneous view of popular science.

Ours is a fairly limited case study, pertaining to one book in particular about the Bohr theory of structure of the atom. We have no knowledge of what Bohr himself made of this book, if he was involved, or in some other way promoted Kramers and Holst's efforts. Working closely with Kramers, he must have known about the book, the close collaboration probably being one of the reasons why Bohr's own thoughts regarding the book project remains unknown: if it ever entered into their exchanges (as it undoubtedly did), Kramers and Bohr most likely would have simply talked about it. We do know that Bohr, despite limited resources in this respect, regarded popularisation a valuable exercise in itself. He believed that popular science involved more than simple translations of scientific knowledge; to him, popularisation was more like an extension of the scientific endeavour to make physical concepts and ideas as clear as possible. If popularisation efforts failed, then it might as well indicate that the original science suffered from certain flaws.

Except from the range of books surveyed and a few reviews of Kramers and Holst's book in scientific journals, we have very limited knowledge about popular receptions of the Bohr theory in contemporary culture. There still is important historical work to be done in locating newspaper articles, visual imagery, jokes, letters etc., all referring to quantum theory in the period. There is little doubt that, by the early 1920's, atomic models had become a matter of sustained public interest; yet, we have little knowledge about this process of 'enculturation' of the quantum atom.

[87] Stephen Hilgartner, 'The Dominant Model of Popularization: Conceptual Problems, Political Uses', *Social Studies of Science*, 20 (1990), 519-539. D. Paul, 'Spreading Chaos: The Role of Popularizations in the Diffusion of Scientific Ideas', *Written Communication*, 21 (2004), 32-68.

Our study almost exclusively has dealt with the kind of popular science that emerges from the community of scientists and closely affiliated science writers, and we remain convinced that the popular view of the Bohr atom that has emerged from our material will be somewhat modified by future studies.

The question concerning the impact of the Bohr atom on science and on visual culture aspires to become one of the fruitful avenues for future historical research. Arne Schirrmacher, Charlotte Bigg, Jochen Hennig and coworkers already have laid the foundations in their studies of a wide range of atomic images (*Atombilder*).[88] We found that the pictorial atoms included in Kramers and Holst's book, despite the fact that they never were intended to portray real atoms, travelled particularly well across national boundaries, across the science-popular science divide, and across time. In public, even today, the planetary model of the atom has become ubiquitous, partly in consequence of its adoption for the Atoms for Peace Campaign of the Eisenhower Administration of the 1950s.[89] Admittedly, there is huge historical gap between the popularisation of the Bohr theory using images in the 1920s and the political war waged by the USA during the Cold War period. However, pictorial atoms appear to be rather resistant to changes in historical context, and we, as historians of science, have only begun to understand why this is so.

Notwithstanding that our knowledge of the popularisation of the Bohr atom is still incomplete, we still would like to suggest a few general conclusions about the role of popular physics in our period. Certain aspects of the popularisation of the Bohr theory are easily accommodated into the standard model of an increasing gap between science and the public.[90] In its scientific articulation, the Bohr theory was highly abstract and mathematically complicated. Popular renditions, therefore, used everyday language and metaphors as well as visual imagery to translate the theory into popular parlance. In so doing, popular physics widened the gap between scientific knowledge and popular opinion.

---

[88] Bigg and Hennig (note 8).
[89] Kenneth Osgood, *Total Cold War: Eisenhower's Secret Propaganda Battle at Home and Abroad* (Lawrence, KS: University Press of Kansas, 2006), chapter 5.

However, we also find that certain aspects of our story suggest more fluid boundaries between science and popular science. Popular science, surely, was used to disseminate new knowledge to various audiences, but this did not entail fixing the limits between scientific knowledge in-the-making and ready-made science.[91] Some scientists were active as popularisers without having a clear notion of when to do what. Arnold Sommerfeld, for example, wanted to give a popular exposition of quantum theory in his *Atombau*, but ended up writing a widely used textbook for physicists. Niels Bohr, particularly so in his later days, understood everyday language as fundamental even to scientific explanations of atomic phenomena Moreover, physicists read and reviewed popular books about quantum theory. More than simply neutral accounts, such books provided feedback into scientific discussions and inspired new physicists to take up quantum theory. One such physicist was Victor Weisskopf, who recalled that while at high school, at the age of 15: 'I was also influenced by reading an enormous amount of popular literature on Bohr; in fact the 10th anniversary issue of *Naturwissenschaften* I read word by word, not understanding very much. … I also read Kramers-Holst's book, and the like.'[92]

---

[90] Bensaude-Vincent (note 3).

[91] Using Bruno Latour's terms, see Bruno Latour, *Science in Action: Following Scientists and Engineers Through Society* (Cambridge, Mass.: Harvard University Press, 1987), p. 4.

[92] Interview with Victor F. Weisskopf by T. S. Kuhn and J. L. Heilbron 10 July 1965, Niels Bohr Library & Archives, American Institute of Physics, Washington, <http://www.aip.org/history/ohilist/4944.html> [accessed 6 September 2011].

The boundary between science and popular science is semipermeable, to use a biological metaphor, with knowledge, interests and power flowing back and forth.

*Acknowledgments*:  We would like to thank Finn Aaserud and Felicity Pors at the Niels Bohr Archive, Copenhagen University, for their interest in this study and help with locating sources relevant for it.